\documentclass[12pt]{iopart}

\usepackage{amssymb}
\usepackage{amsthm}
\usepackage{bbm}
\usepackage{graphicx}

\usepackage{color}

\newtheorem{theorem}{Theorem}

\newcommand{\Ket}[1]{\left\vert#1\right\rangle}
\newcommand{\BraKet}[2]{\left\langle#1\right\vert\left.#2\right\rangle}

\newcommand{\h}{\mathcal{H}}
\newcommand{\Ignore}[1]{ }

\begin{document}

\title{An algebraic approach to the study of multipartite entanglement}

\author{S. Di Martino}
\address{Dipartimento di Matematica dell'Universit\`{a} di Bari, Via Orabona 4, 70125 Bari, Italy}

\author{B. Militello}
\address{Dipartimento di Fisica dell'Universit\`{a} di Palermo, Via Archirafi 36, 90123 Palermo, Italy}

\author{A. Messina}
\address{Dipartimento di Fisica dell'Universit\`{a} di Palermo, Via Archirafi 36, 90123 Palermo, Italy}

\begin{abstract}%
A simple algebraic approach to the study of multipartite
entanglement for pure states is introduced together with a class
of suitable functionals able to detect entanglement. On this
basis, some known results are reproduced. Indeed, by investigating
the properties of the introduced functionals, it is shown that a
subset of such class is strictly connected to the purity.
Moreover, a direct and basic solution to the problem of the
simultaneous maximization of three appropriate functionals for
three-qubit states is provided, confirming that the simultaneous
maximization of the entanglement for all possible bipartitions is
compatible only with the structure of GHZ-states.
%
%
%
%
%
\end{abstract}%

\maketitle

\section{Introduction}

Entanglement plays an important role in applications of quantum
mechanics to nanotechnologies, especially in the field of quantum
information and communication. The concept of entanglement is
mathematical, since it corresponds, in the case of pure states, to
the impossibility of writing the quantum state describing a
compound physical system as a simple product of states related to
the single subsystems. Nevertheless, this mathematical property of
the state has interesting physical implications, such as the
presence of non classical correlations between quantum systems,
even when they are quite far from each other (non locality).

In spite of the strong interest and deep studies developed on this
subject, detection and classification of entanglement in
multipartite systems are unsolved problems up to date. Indeed, it
has not yet been defined a universal quantity able to measure the
entanglement level to be associated to a generic pure or even
mixed state. (The concept of non entangled state is translated
into factorizability for pure states and separability
--- simple or generalized --- for mixed states, though the latter
concept includes the former as a special case.) Besides, there are
properties that any measurement of entanglement should possess. In
particular, it must be discriminant (vanishing iff the state is
separable), not increasing under LOCC (local operations and
classical communication) and convex (it must not increase when two
or more states are combined in a mixture)~\cite{ref:Vedral1997}.

Though a general answer to measurement and detection of
entanglement is absent, if the system is composed by a few of low
dimensional subsystems it is possible to provide conditions for
the presence of entanglement, but rarely these conditions are both
necessary and sufficient. In 1996, Peres~\cite{ref:p96} introduced
a sufficient condition for the detection of bipartite entanglement
that can be written in terms of a functional addressed as {\it
negativity}, subsequently studied by Horodecki {\it et al}, who
have proven that this condition is necessary and sufficient for
low-dimensional systems~\cite{ref:Horodecki1997, ref:ho}. Another
important functional is the {\it concurrence} introduced by
Wootters in \cite{ref:hw, ref:w98}, which is a true measure of
entanglement, both for pure and mixed states, but unfortunately
applicable only to systems which are couples of two level systems
(i.e., a couple of {\it qubits}).

The study of multipartite entanglement is more
complex~\cite{ref:re, ref:man, ref:pf}, not only from the
computational point of view, but even at a conceptual level, since
for instance it is neither immediate nor intuitive to understand
what is a state with a maximum level of multipartite entanglement.
In 2008, Sabin {\it et al}, \cite{ref:ga}, tried to extend the
negativity to the tripartite case, succeeding in finding a
functional that detects the tripartite entanglement when applied
to pure states. However, this functional gives only a clue about
the presence of entanglement in mixed states, indeed in this case
the condition given is neither necessary nor sufficient.
Nevertheless, its effectiveness has been shown in the study of
simple physical systems~\cite{ref:fa}. Another interesting
quantity, introduced for detecting entanglement of pure states of
three qubits, is the three-tangle~\cite{ref:tt}, which is based on
the concurrence and whose validity has been
criticized~\cite{ref:ttt}. Recently, a classification of entangled
states for three-qubit states has been given on the basis of some
invariant quantities \cite{ref:classification}. Moreover, attempts
to apply well known results of algebra or geometry to study
multipartite entanglement have been made. In \cite{ref:mm}, for
instance, M\"{a}kel\"{a} {\it et al} associated to every pure
state of N-qubits a polynomial, characterizing completely the
state, capable not only to detect factorizability of the state but
even the number of separable qubits. Instead, in \cite{ref:hyp},
Miyake used the hyperdeterminants and the theory of Segr\`{e}
variety to give a classification of the entanglement of pure
states in tripartite systems. A very useful tool to study
multipartite entanglement in mixed states is given by the
necessary condition for separability expressed by Huber {\it et
al}~\cite{ref:hu}, which exploits a correlation function defined
through replicas of the states under scrutiny and operators of
partial swapping, and which has been exploited to reveal thermal
tripartite correlations in spin-star systems~\cite{ref:te}.

Another important quantity that has been used to reveal
entanglement in several cases (bipartite and multipartite
systems), is the purity (strictly connected to the linear
entropy~\cite{ref:Fano1957}) of the reduced density matrix, which
in passing is strictly connected with the concept of
mixedness~\cite{ref:Jaeger2003}: the more two systems are
entangled, the less pure is the reduced state describing each one
of the two systems.

In this paper we reproduce some known results about entanglement
detection by introducing a simple algebraic approach to establish
whether a pure state of a multipartite system is entangled with
respect to a given bipartition. Through this analysis it is
possible to determine whether a given pure state is completely
separable, separable or totally entangled. In the last case one
can infer the presence of genuine multipartite entanglement. On
the basis of this approach, we naturally introduce a class of
functionals which includes, among others, quantities traceable
back to the purity and the linear entropy, strictly related to the
so called concurrence vectors, presented in~\cite{ref:osterloh} by
Akhtarshenas and studied by Mintert {\it et al}
in~\cite{ref:huber1,ref:huber2}.

%

The paper is organized as follows. In the next section we describe
our approach to the study of multipartite entanglement and give
the definition of a relevant class functionals, proving the
ability of such quantities to detect entanglement in pure states.
In section \ref{sec:Demitization} we show that such quantities are
strictly connected with purity and mixedness. In section
\ref{sec:Maximum} we show, by exploiting our simple and algebraic
approach, that in the case of three qubits, simultaneous
maximization of the three relevant functionals is possible only
for states equivalent (i.e., equal up to a local and unitary
transformation) to the GHZ state. Finally, in the last section, we
give some comments and concluding remarks.

\section{Factorizability conditions} \label{sec:fact_cond}

Each pure state of a bipartite system can be written in the form:

\begin{equation}
\Ket{\phi}=\sum_{i,j}a_{ij}\Ket{ij},
\end{equation}
where we take $\{\Ket{ij}\}$ as a standard basis in the Hilbert
space $\h=\h_1\otimes\h_2$. This state is said factorizable if it
can be written in the form:

\begin{equation}
(\sum_i\alpha_i\Ket{i})(\sum_j\beta_j\Ket{j}).
\end{equation}

This means that the state is factorizable iff
$a_{ij}=\alpha_i\beta_j,$ $\forall i,j$. So, introducing the
matrix of the probability amplitudes $A=(a_{ij})$, the state
$\Ket{\phi}$ is factorizable iff columns (respectively rows)
span a $1$-dimensional subspace, i.e. $rank A=1$.

In general, this means $\det A=0$ and in the case of a two qubits system, in
\cite{ref:Equiv_neg} it is proven that this condition is
equivalent to the Peres-Horodecki criterion. The condition here
expressed can be easily extended to investigate the presence of
entanglement in every pure states of multipartite systems. We can
summarize this result for bipartite systems with the following
theorem.

\begin{theorem}
A pure state of a bipartite $m\times n$ system is factorizable iff
the rank of the corresponding matrix of the probability amplitudes
is unity.
\end{theorem}

\bigskip

Now we can extend this result to tripartite systems and we can
generalize it to every multipartite system. Consider a $m\times
n\times p$ Hilbert space. Each pure state $\Ket{\phi}$ living in
it can be written in the form:
\begin{equation}
|\phi\rangle=\sum_{i=0}^{m-1}\sum_{j=0}^{n-1}\sum_{k=0}^{p-1}a_{ijk}|ijk\rangle,
\end{equation}
where $\{\Ket{ijk}\}$ is the standard product basis for the
Hilbert space related to the whole system. Again, the state is factorizable with respect, for instance, to the first
component, iff:
\begin{equation}
a_{ijk}=\alpha_i\beta_{jk}\, \qquad \forall i,j,k.
\label{sep_cond}
\end{equation}

This time the matrix of the probability amplitudes is a cubic one
and the separability condition of the first component can be
traced back to the proportionality of the layers in the direction
of $i$. For a better visualization of this statement we can write
two layers in the $i$-direction:
\begin{center}
\mbox{ $\left( \begin{array}{ccc}
\alpha_r\beta_{11}&\cdots&\alpha_r\beta_{1p}\\
\cdots&&\\
\alpha_r\beta_{n1}&\cdots&\alpha_r\beta_{np} \\
\end{array} \right)$
$\quad$,$\quad$ $\left(\begin{array}{ccc}
\alpha_{s}\beta_{11}&\cdots&\alpha_{s}\beta_{1p}\\
\cdots&&\\
\alpha_s\beta_{n1}&\cdots&\alpha_{s}\beta_{np} \\
\end{array}\right)$,}
\end{center}
which are the $r^{th}$ and $s^{th}$ layers.

Starting from these considerations, we can prove the following
theorem.

\begin{theorem}
\label{trip1} A pure state $\Ket{\phi}$ in a $m\times n\times p$
Hilbert space is factorizable iff one of the following is true:

\begin{itemize}
\item $a_{ijk}a_{i'j'k'}-a_{ij'k'}a_{i'jk}=0\,,\qquad \forall i,j,k,i',j',k'$;
\item $a_{ijk}a_{i'j'k'}-a_{i'jk'}a_{ij'k}=0\,,\qquad \forall i,j,k,i',j',k'$;
\item $a_{ijk}a_{i'j'k'}-a_{i'j'k}a_{ijk'}=0\,,\qquad \forall i,j,k,i',j',k'$.
\end{itemize}

\end{theorem}
\proof For the sake of simplicity and without loss of generality,
let us consider the factorizability condition for the first qubit.

If the state $\Ket{\phi}$ is factorizable with respect to such
first qubit, using the factorizability condition in equation
(\ref{sep_cond}) we have:

\begin{equation}
\alpha_i\beta_{jk}\alpha_{i'}\beta_{j'k'}-\alpha_i\beta_{j'k'}\alpha_{i'}\beta_{jk}=0.
\end{equation}
Vice versa, if we suppose $a_{ijk}a_{i'j'k'}-a_{ij'k'}a_{i'jk}=0, \forall i,j,k,i',j',k'$ then, fixing $i,j,k$ such that $a_{ijk}\neq 0$, it turns out that:

\begin{equation}
a_{i'j'k'}=\frac{a_{i'jk}}{a_{ijk}}a_{ij'k'}=\alpha_{i'}\beta_{j'k'}\,,
\,\,\,\,\, \forall i',j',k',
\end{equation}
where we take $\alpha_{i'}=\frac{a_{i'jk}}{a_{ijk}}$ and
$\beta_{j'k'}=a_{ij'k'}$, which is the factorizability condition
of the first component.
\endproof

Using this theorem we can define a class of functionals able to
detect entanglement in each pure state of a tripartite system.

Let $f:\mathbb{C}\rightarrow \mathbb{R} $ be a positive function
such that $\forall x\in \mathbb{C}\, f(x)\geq 0$ and $f(x)=0$ iff
$x=0$, then we have the following:

\begin{theorem}
\label{trip} A pure state $\Ket{\phi}$ in a $m\times n\times p$
Hilbert space is factorizable iff at least one of the following is
true:
\begin{center}
\begin{itemize}
\item $M_1^{(f)}:=\sum_{i,j,k,i',j',k'}f(a_{ijk}a_{i'j'k'}-a_{ij'k'}a_{i'jk})=0;$
\item $M_2^{(f)}:=\sum_{i,j,k,i',j',k'}f(a_{ijk}a_{i'j'k'}-a_{i'jk'}a_{ij'k})=0;$
\item $M_3^{(f)}:=\sum_{i,j,k,i',j',k'}f(a_{ijk}a_{i'j'k'}-a_{i'j'k}a_{ijk'})=0.$
\end{itemize}
\end{center}
\end{theorem}

The proof of this result follows immediately from the theorem
\ref{trip1} and from the properties of $f$.

We can make some observation about this. Firstly, if $M_k^{(f)}=0$
then the corresponding state is factorizable with respect to the
$k^{th}$ component, so if two of these quantities are zero also
the third has to be equal to zero (considering that, for pure
states, if two of the three subsystems are separable, then the
third one is separable as well). This statement can be easily
proven considering the factorizability condition in equation
(\ref{sep_cond}) for two of the three components. Secondly, if all
the three $M_k^{(f)}$ are equal to zero than the state is
completely factorizable --- i.e., it is the product of three
states of the three subsystems --- and if each one of them is
different from zero the state is genuinely tripartite entangled.

\bigskip

Theorem \ref{trip} can be extended to the multipartite case
considering all the possible bipartitions of the system. For
instance, in the case of a quadripartite system we have seven
conditions given by:
\begin{center}
\begin{itemize}

\item $M_1^{(f)}:=\sum_{i,j,k,l,i',j',k',l'}f(a_{ijkl}a_{i'j'k'l'}-a_{ij'k'l'}a_{i'jkl})=0;$
\item $M_2^{(f)}:=\sum_{i,j,k,l,i',j',k',l'}f(a_{ijkl}a_{i'j'k'l'}-a_{i'jk'l'}a_{ij'kl})=0;$
\item $M_3^{(f)}:=\sum_{i,j,k,l,i',j',k',l'}f(a_{ijkl}a_{i'j'k'l'}-a_{i'j'kl'}a_{ijk'l})=0;$
\item $M_4^{(f)}:=\sum_{i,j,k,l,i',j',k',l'}f(a_{ijkl}a_{i'j'k'l'}-a_{i'j'k'l}a_{ijkl'})=0;$
\item $M_{12}^{(f)}:=\sum_{i,j,k,l,i',j',k',l'}f(a_{ijkl}a_{i'j'k'l'}-a_{ijk'l'}a_{i'j'kl})=0;$
\item $M_{13}^{(f)}:=\sum_{i,j,k,l,i',j',k',l'}f(a_{ijkl}a_{i'j'k'l'}-a_{ij'kl'}a_{i'jk'l})=0;$
\item $M_{23}^{(f)}:=\sum_{i,j,k,l,i',j',k',l'}f(a_{ijkl}a_{i'j'k'l'}-a_{i'jkl'}a_{ij'k'l})=0.$

\end{itemize}
\end{center}
where $M_k^{(f)}$ functionals are related to separation of the $k$-th
subsystem when the other three are considered as a whole, while
$M_{kj}^{(f)}$'s are related to the bipartition obtained by considering
the couple $k-j$ as a whole and the other two parts as the second
subsystem of the bipartition.

\section{Connection with the Purity}\label{sec:Demitization}

In this section we prove that if we take as function $f$ the
squared modulus, the relevant functionals, simply denoted as
$M_k$, are strictly connected with the purity of the relevant
reduced density operators. To this end, let us first of all
rewrite $M_k$ by considering only two indexes (this means that the
two indexes, say $i$ and $j$, will span multipartite subsystems):
\begin{equation}
M_k=\sum_{i,j,i',j'}|a_{ij}a_{i'j'}-a_{ij'}a_{i'j}|^2.
\end{equation}
Then, by expanding the modulus, one gets:
\begin{eqnarray}
\nonumber
M_k&=&\sum_{i,j,i',j'}(a_{ij}a_{ij}^*a_{i'j'}a_{i'j'}^*+a_{ij'}a_{ij'}^*a_{i'j}a_{i'j}^*-a_{ij}a_{i'j'}a_{ij'}^*a_{i'j}^*-a_{ij}^*a_{i'j'}^*a_{ij'}a_{i'j})\\
&=&2(1-p)
\end{eqnarray}
where
\begin{equation}
p = \sum_{i,j,i',j'} a_{ij}a_{i'j'}a_{ij'}^*a_{i'j}^* =
\sum_{i,j,i',j'} a_{ij}^*a_{i'j'}^*a_{ij'}a_{i'j}\,,
\end{equation}
is the purity of the density operator associated to any of the two
subsystems constituting the bipartition. This makes the functional
$M_k$ essentially equal (up to the factor $2$) to the linear
entropy (mixedness)~\cite{ref:Jaeger2003}.

Therefore, the quantities $M_k$ are invariant under local unitary
transformations and are bounded as follows: $0 \le M_k \le
2(D-1)/D$, where $D$ is the dimension of the smaller among the two
Hilbert subspaces associated to the bipartition considered.
Moreover, maximization is reached when the whole-system state can
be written as a superposition of $D$ states with the same weights:

\begin{equation}
\Ket{\psi}=D^{-1/2}\sum_{j=1}^D e^{i\chi_j}
\Ket{j}_k\Ket{\Phi_j}_{\bar{k}},
\end{equation}
where $\Ket{j}_k$ and $\Ket{\Phi_j}_{\bar{k}}$ refer to the two
subsystems constituting the bipartition, and
$\BraKet{j}{l}=\BraKet{\Phi_j}{\Phi_l}=\delta_{jl}$.

\subsection{Some Examples}

\begin{figure}
\centering
\includegraphics[width=0.40\textwidth, angle=0]{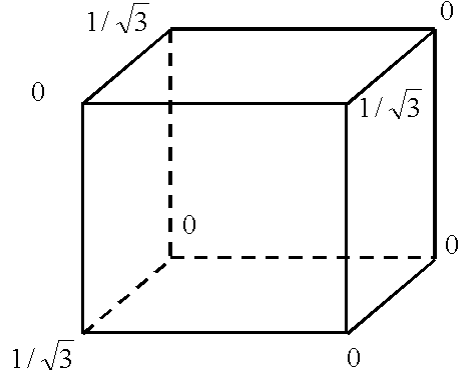}
\caption{Matrix of the probability amplitudes related to the
W-state. The vertexes of the first layer correspond (from top-part
left-side, clockwise) to the coefficients: $a_{000}$, $a_{001}$,
$a_{011}$, $a_{010}$; the vertexes of the second layer instead
correspond to the coefficients: $a_{100}$, $a_{101}$, $a_{111}$,
$a_{110}$.} \label{cuboW}
\end{figure}

For a better understanding of the results given in the previous
section we apply the previously proven criteria to some pure
states. First of all, let us consider the W-state which, as well
known, is genuinely tripartite entangled:
\begin{equation}
|W\rangle=\frac{1}{\sqrt{3}}(|100\rangle+|010\rangle+|001\rangle).
\end{equation}
The relevant cubic matrix of probability amplitudes is shown in
fig.\ref{cuboW}.

A rapid calculation gives:
\begin{equation}
M_1=M_2=M_3=\frac{8}{9},
\end{equation}
and therefore, the state is genuinely tripartite entangled in
accordance with theorem \ref{trip}.

\bigskip

\begin{figure}
\centering
\includegraphics[width=0.40\textwidth, angle=0]{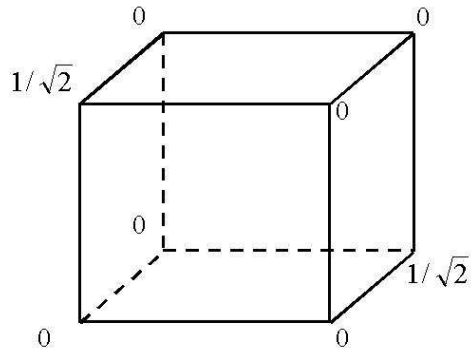}
\caption{Matrix of the probability amplitudes related to the
GHZ-state. The vertex-coefficient correspondence is the same as in
the previous figure.} \label{cuboGHZ}
\end{figure}

As a second example, let us consider the GHZ-state:
\begin{equation}
|GHZ\rangle=\frac{1}{\sqrt{2}}(|000\rangle+|111\rangle),
\end{equation}
whose matrix of the probability amplitudes is shown in fig.
\ref{cuboGHZ}.

\newpage

Once again a straightforward calculation leads to:
\begin{equation}
M_1=M_2=M_3=1.
\end{equation}

By the way, we mention that in the following section we prove
unity to be the maximum value that these quantities can reach for
a three-qubit system. Moreover, it is worth noting that
maximization of two $M_k$ does not imply maximization of the
third, as shown by the following example. Indeed, consider the
state:
\begin{equation}
|\psi_{bis}\rangle=\frac{1}{\sqrt{2}}\Ket{000}+\frac{1}{\sqrt{2}}\Ket{011},
\end{equation}
which is factorizable with respect to the first qubit, being
$M_1=0$, but it is not factorizable with respect to the other two
qubits, being $M_2=M_3=1$.

\bigskip

Finally, let us consider a state for which it is not evident
whether it is entangled or not:
\begin{equation}
\Ket{\varphi}=\frac{1}{2\sqrt{26}}(5\Ket{000}+3\Ket{010}+2\Ket{001}+4\Ket{011}+7\Ket{101}+\Ket{111}),
\end{equation}
Since it turns out that $M_1\approx 0.88$, $M_2\approx 0.42$ and
$M_3\approx 0.70$
we can conclude that it is a genuinely tripartite entangled state.

\section{Maximization of $M$-quantities for a three qubit
system}\label{sec:Maximum}

In the second example of section \ref{sec:fact_cond} we proved that
for the GHZ-state all the $M_k$'s reach their maximum value. In this
section we prove that all pure states of three qubits for which
$M_k=1\, \forall k$ are equivalent to the GHZ-state, i.e. there
exists a local and unitary transformation that maps this state into
GHZ-state.


Our simple and basic approach makes more transparent the proof of
this statement in comparison with the proof given in
\cite{ref:GHZ}.

\bigskip

Consider a three qubits pure state $\Ket{\psi}$ for which it is valid
the hypothesis $M_k=1\, \forall k$. Since for this state the $M_k$
reach their maximum, we can write the state in the following form:

\begin{equation}
\Ket{\psi}=\frac{1}{\sqrt{2}} \Ket{0}_k\Ket{\phi_0}_{\bar{k}}+\frac{1}{\sqrt{2}}e^{i\chi} \Ket{1}_k\Ket{\phi_1}_{\bar{k}},
\end{equation}
where $\bar{k}$ refers to the two indexes different from $k$,
$\{\Ket{0},\Ket{1}\}$ is the standard basis of the $k^{th}$ qubit
and $\Ket{\phi_0}$ and $\Ket{\phi_1}$ are such that
$\BraKet{\phi_a}{\phi_b}=\delta_{ab}$ for $a,b=0,1$, simply
applying a Schmidt's decomposition.

By the way, we note that in the case of two qubits, maximization of
the unique $M$-functional leads to a Bell-like structure.

In the tripartite case, because of the simultaneous maximization
of the three $M_k$'s, we have that this structure is valid for each
of the three qubits.

Moreover, this structure is left unchanged by any unitary and
local transformation.

Let us consider this structure with respect to the first qubit
and, applying again a Schmidt's decomposition, we can convert the
state $\Ket{\phi_0}_{23}$ in
$(\cos\theta_1\Ket{00}_{23}+e^{i\chi_1}\sin\theta_1\Ket{11}_{23})$,
getting:

\begin{eqnarray}
\label{pq} \nonumber
\Ket{\psi}=&\frac{1}{\sqrt{2}} \Ket{0}_1(\cos\theta_1\Ket{00}_{23}+e^{i\chi_1}\sin\theta_1\Ket{11}_{23})\\
&+\frac{1}{\sqrt{2}}e^{i\chi} \Ket{1}_1[\Ket{0}_2(a\Ket{0}_3+b\Ket{1}_3)+\Ket{1}_2(c\Ket{0}_3+d\Ket{1}_3)],
\end{eqnarray}
where $\{\Ket{0}_l, \Ket{1}_l\}$ is a basis, not
necessarily standard, for the $l^{th}$ qubit, and $a,b,c,d$ are
complex numbers and every coefficient satisfies the normalization
and orthogonality condition between the two states
$\Ket{\phi_0}_{23}$ and $\Ket{\phi_1}_{23}$.

Now we can decompose the state $\Ket{\psi}$ with respect to the
second qubit, getting:

\begin{eqnarray}
\nonumber
\Ket{\psi}=&\frac{1}{\sqrt{2}} \Ket{0}_2[\cos\theta_1\Ket{00}_{13}+e^{i\chi}\Ket{1}_1(a\Ket{0}_3+b\Ket{1}_3)]\\
&+\frac{1}{\sqrt{2}}\Ket{1}_2[e^{i\chi_1}\sin\theta_1\Ket{01}_{13}+e^{i\chi}\Ket{1}_1(c\Ket{0}_3+d\Ket{1}_3)].
\end{eqnarray}

The orthogonality conditions involves the orthogonality between
the two states $(a\Ket{0}_3+b\Ket{1}_3)$ and
$(c\Ket{0}_3+d\Ket{1}_3)$, which implies $a=\alpha\cos\theta_2$, $b=\alpha e^{i\chi_2}\sin\theta_2$, $c=\beta e^{-i\chi_2}\sin\theta_2$ and $d=\beta\cos\theta_2$.

In addition, exploiting the
normalization constrain with respect to the first qubit, we find $\alpha=\cos\theta_3$ and $\beta=e^{i\chi_3}\sin\theta_3$, so that we can rewrite the state in equation (\ref{pq}) as:

\begin{equation}
\Ket{\psi}=\frac{1}{\sqrt{2}} \Ket{0}_1\Ket{\phi_0}_{23}+\frac{1}{\sqrt{2}}e^{i\chi}\Ket{1}_1\Ket{\phi_1}_{23},
\end{equation}
where

\begin{eqnarray}
\nonumber
\Ket{\phi_0}_{23}=(\cos\theta_1\Ket{00}_{23}+e^{i\chi_1}\sin\theta_1\Ket{11}_{23})
\end{eqnarray}
and

\begin{eqnarray}
\nonumber
\Ket{\phi_1}_{23}=&\cos\theta_3\Ket{0}_2(\cos\theta_2\Ket{0}_3+e^{i\chi_2}\sin\theta_2\Ket{1}_3\\
\nonumber
&+e^{i\chi_3}\sin\theta_3\Ket{1}_2(e^{-i\chi_2}\sin\theta_2\Ket{0}_3-\cos\theta_2\Ket{1}_3).
\end{eqnarray}

Summarizing, we impose the normalization condition of
$\Ket{\phi_0}$ e $\Ket{\phi_1}$ with respect to the first qubit
and the orthogonality condition with respect to the second qubit.
We now have to enforce: 1)The orthogonality condition with respect
to the first qubit; 2)The normalization condition with respect to
the second qubit; 3)The orthogonality condition with respect to
the third qubit; 4)The normalization condition with respect to the
third qubit.

The orthogonality condition for the first qubit leads to $e^{i\chi}\cos\theta_1\cos\theta_2\cos\theta_3-e^{i(\chi+\chi_3-\chi_1)}\sin\theta_1\cos\theta_2\sin\theta_3=0$, and so we have the condition:

\begin{equation}
\label{c1}
\theta_2=\frac{\pi}{2} \quad \vee \quad \cos\theta_1\cos\theta_3-e^{i(\chi_3-\chi_1)}\sin\theta_1\sin\theta_3=0.
\end{equation}

Rewriting the state $\Ket{\psi}$ as,

\begin{eqnarray}
\nonumber
\Ket{\psi}=&\frac{1}{\sqrt{2}} \Ket{0}_2[\cos\theta_1\Ket{00}_{13}+e^{i\chi}\cos\theta_3\Ket{1}_1(\cos\theta_2\Ket{0}_3+e^{i\chi_2}\sin\theta_2\Ket{1}_3)]\\
&+\frac{1}{\sqrt{2}}\Ket{1}_2[\sin\theta_1\Ket{01}_{13}\\
\nonumber
&+e^{i(\chi+\chi_3)}\sin\theta_3\Ket{1}_1(e^{-i\chi_2}\sin\theta_2\Ket{0}_3-\cos\theta_2\Ket{1}_3)],
\end{eqnarray}
the normalization condition for the second qubit leads to
$(\cos^2\theta_1+\cos^2\theta_3\cos^2\theta_2+\cos^2\theta_3\sin^2\theta_2=1)$,
that is,

\begin{equation}
\label{c2}
\cos\theta_3=\sin\theta_1,
\end{equation}
and likewise for the other state (we consider only this
case because possible sign difference between sine and
cosine can be described by adjusting the phase factors).

Now, we decompose the state with respect to the third qubit:

\begin{equation}
\Ket{\psi}=\frac{1}{\sqrt{2}}\Ket{0}_3\Ket{\phi '_0}_{12}+\frac{1}{\sqrt{2}}\Ket{1}_3\Ket{\phi '_1}_{12},
\end{equation}
where

\begin{eqnarray}
\nonumber
\Ket{\phi '_0}_{12}=&(\cos\theta_1\Ket{00}_{12}+e^{i\chi}\cos\theta_3\cos\theta_2\Ket{10}_{12}\\
\nonumber
&+e^{i(\chi+\chi_3-\chi_2)}\sin\theta_3\sin\theta_2\Ket{11}_{12})
\end{eqnarray}
and
\begin{eqnarray}
\nonumber
\Ket{\phi '_1}_{12}=&&(e^{i\chi_1}\sin\theta_1\Ket{01}_{12}+e^{i(\chi+\chi_2)}\cos\theta_3\sin\theta_2\Ket{10}_{12}\\
\nonumber
&&-e^{i(\chi+\chi_3)}\sin\theta_3\cos\theta_2\Ket{11}_{12}),
\end{eqnarray}
and we impose the third condition (orthogonality with respect to
the third qubit):
$(e^{i\chi_2}\cos^\theta_3\cos\theta_2\sin\theta_2-e^{i\chi_2}\sin^2\theta_3\cos\theta_2\sin\theta_2=0)$,
that leads to

\begin{equation}
\label{c3}
\theta_3=\frac{\pi}{4} \quad \vee \quad \theta_2=0 \quad \vee \quad \theta_2=\frac{\pi}{2},
\end{equation}
(again we consider only $\theta_3=\frac{\pi}{4}$ because possible
sign difference between $\sin\theta_3$ and $\cos\theta_3$ can be
englobed in the phase factors).

Summarizing, we can have three
possible situations:

\begin{equation}
\begin{array}{c}
\theta_3=\frac{\pi}{4}\Rightarrow\theta_1=\frac{\pi}{4};\\
\\
\theta_2=0;\\
\\
\theta_2=\frac{\pi}{2}.\\
\end{array}
\end{equation}

The normalization condition at point $4$,
$\cos^2\theta_1+\cos^2\theta_3\cos^2\theta_2+\sin^2\theta_3\sin^2\theta_2=1$
is automatically satisfied for $\theta_3=\frac{\pi}{4}$ and
$\theta_1=\frac{\pi}{4}$ as for $\theta_2=0$. Instead, if
$\theta_2=\frac{\pi}{2}$ then necessarily
$\theta_1=\theta_3=\frac{\pi}{4}$.

We can apply the conditions we have found, writing the initial
state as:

\begin{eqnarray}
\nonumber
&\frac{1}{\sqrt{2}}\Ket{0}_1(\sin\theta_3\Ket{00}_{23}+e^{i\chi_1}\cos\theta_3\Ket{11}_{23})\\
&+\frac{1}{\sqrt{2}}e^{i\chi}\Ket{1}_1[\cos\theta_3\Ket{0}_2(\cos\theta_2\Ket{0}_3+e^{i\chi_2}\sin\theta_2\Ket{1}_3)\\
\nonumber
&+e^{i\chi_3}\sin\theta_3(e^{-i\chi_2}\sin\theta_2\Ket{0}_3-\cos\theta_2\Ket{1}_3)],
\end{eqnarray}
where parameters vary in the following way:

\begin{equation}
\label{tab}
\begin{array}{c}
\theta_3=\frac{\pi}{4}\quad \forall \theta_2;\\
\\
\theta_2=0\quad \forall \theta_3;\\
\end{array}
\end{equation}

It remains to prove that this state is equivalent to the GHZ under
local unitary transformation.

If $\theta_2=0$, condition (\ref{c1}) assures either
$\chi_1=\chi_3$ or $\theta_3=0$ or $\theta_3=\frac{\pi}{2}$. Let
us examine these three possibilities.

If $\theta_2=0$ and $\theta_3=0$ the
state becomes:

\begin{equation}
\frac{1}{\sqrt{2}}e^{i\chi_3}\Ket{011}+\frac{1}{\sqrt{2}}e^{i\chi}\Ket{100},
\end{equation}
and we can easily find a transformation that sends it in the GHZ.

If $\theta_2=0$ and $\theta_3=\frac{\pi}{2}$ then we have the state:

\begin{equation}
\frac{1}{\sqrt{2}}\Ket{000}-\frac{1}{\sqrt{2}}e^{i(\chi+\chi_3)}\Ket{111},
\end{equation}
and even in this case we can easily find the transformation needed.

If, finally, $\theta_2=0$ and $\chi_1=\chi_3$ the state becomes:

\begin{eqnarray}
\nonumber
&\frac{1}{\sqrt{2}}\Ket{0}_1(\sin\theta_3\Ket{00}_{23}+e^{i\chi_3}\cos\theta_3\Ket{11}_{23})\\
\nonumber
&+\frac{1}{\sqrt{2}}e^{i\chi}\Ket{1}_1(\cos\theta_3\Ket{00}_{23}-e^{i\chi_3}\sin\theta_3\Ket{11}_{23}),\\
\end{eqnarray}
and it is converted in the GHZ by:

\begin{equation}
\begin{array}{rcl}
\sin\theta_3\Ket{0}_1+e^{i\chi}\cos\theta_3\Ket{1}_1&\rightarrow&\Ket{\widetilde{0}}\\
e^{i\chi_3}\cos\theta_3\Ket{0}_1-e^{i(\chi+\chi_3)}\sin\theta_3\Ket{1}_1&\rightarrow&\Ket{\widetilde{1}}\\
\end{array}
\end{equation}

\bigskip

We have now to examine the first case of eq.(\ref{tab}), which is $\theta_3=\frac{\pi}{4}$. In this analysis we have to distinguish the two cases $\theta_2=\frac{\pi}{2}$ and $\theta_2\neq\frac{\pi}{2}$.

If $\theta_3=\frac{\pi}{4}$ and $\theta_2\neq\frac{\pi}{2}$, the state has the form:

\begin{eqnarray}
\nonumber
&\frac{1}{\sqrt{2}}\Ket{0}_1\left(\frac{1}{\sqrt{2}}\Ket{00}_{23}+\frac{1}{\sqrt{2}}e^{i\chi_3}\Ket{11}_{23}\right)\\
&+\frac{1}{\sqrt{2}}\Ket{1}_1[\frac{1}{\sqrt{2}}\Ket{0}_2(\cos\theta_2\Ket{0}_3+e^{i\chi_2}\sin\theta_2\Ket{1}_3)\\
\nonumber
&+\frac{1}{\sqrt{2}}e^{i\chi_3}\Ket{1}_2(e^{-i\chi_2}\sin\theta_2\Ket{0}_3-\cos\theta_2\Ket{1}_3)],
\end{eqnarray}
(remember the (\ref{c1}) assures $\chi_1=\chi_3$).

In this situation the transformation:

\begin{equation}
\begin{array}{rcl}
\Ket{0}_1&\rightarrow&\frac{1}{\sqrt{2}}\Ket{\widetilde{0}}_1+\frac{1}{\sqrt{2}}e^{i\chi}\Ket{\widetilde{1}}_1\\
\\
\Ket{1}_1&\rightarrow&\frac{1}{\sqrt{2}}\Ket{\widetilde{0}}_1-\frac{1}{\sqrt{2}}e^{i\chi}\Ket{\widetilde{1}}_1\\
\\
\Ket{0}_2&\rightarrow&\cos\frac{\theta_2}{2}\Ket{\widetilde{0}}_2+e^{i(\chi_1-\chi_2)}\sin\frac{\theta_2}{2}\Ket{\widetilde{1}}_2\\
\\
\Ket{1}_2&\rightarrow&\sin\frac{\theta_2}{2}\Ket{\widetilde{0}}_2-e^{i(\chi_1-\chi_2)}\cos\frac{\theta_2}{2}\Ket{\widetilde{1}}_2\\
\\
\Ket{0}_3&\rightarrow&\cos\frac{\theta_2}{2}\Ket{\widetilde{0}}_3+e^{i\chi_2}\sin\frac{\theta_2}{2}\Ket{\widetilde{1}}_3\\
\\
\Ket{1}_3&\rightarrow&\sin\frac{\theta_2}{2}\Ket{\widetilde{0}}_3-e^{i\chi_2}\cos\frac{\theta_2}{2}\Ket{\widetilde{1}}_3\\
\end{array}
\end{equation}
maps the GHZ-state in the one found.

If $\theta_3=\frac{\pi}{4}$ and  $\theta_2=\frac{\pi}{2}$ then the
state reduces to:

\begin{eqnarray}
\nonumber
&\frac{1}{\sqrt{2}}\Ket{0}\left(\frac{1}{\sqrt{2}}\Ket{00}+e^{i\chi_1}\frac{1}{\sqrt{2}}\Ket{11}\right)\\
&+\frac{1}{\sqrt{2}}e^{i\chi}\Ket{1}\left(e^{i\chi_2}\frac{1}{\sqrt{2}}\Ket{01}+e^{i(\chi_3-\chi_2)}\frac{1}{\sqrt{2}}\Ket{10}\right),
\end{eqnarray}
and we can apply the transformation:

\begin{equation}
\begin{array}{rcl}
\Ket{0}_1&\rightarrow&\frac{1}{\sqrt{2}}\Ket{\widetilde{0}}_1+\frac{1}{\sqrt{2}}e^{i(\chi-\frac{\chi_1}{2}-\frac{\chi_3}{2})}\Ket{\widetilde{1}}_1\\
\\
\Ket{1}_1&\rightarrow&\frac{1}{\sqrt{2}}\Ket{\widetilde{0}}_1-\frac{1}{\sqrt{2}}e^{i(\chi-\frac{\chi_1}{2}-\frac{\chi_3}{2})}\Ket{\widetilde{1}}_1\\
\\
\Ket{0}_2&\rightarrow&\frac{1}{\sqrt{2}}\Ket{\widetilde{0}}_2+\frac{1}{\sqrt{2}}e^{i(\frac{\chi_1}{2}-\chi_2+\frac{\chi_3}{2})}\Ket{\widetilde{1}}_2\\
\\
\Ket{1}_2&\rightarrow&\frac{1}{\sqrt{2}}\Ket{\widetilde{0}}_2-\frac{1}{\sqrt{2}}e^{i(\frac{\chi_1}{2}-\chi_2+\frac{\chi_3}{2})}\Ket{\widetilde{1}}_2\\
\\
\Ket{0}_3&\rightarrow&\frac{1}{\sqrt{2}}\Ket{\widetilde{0}}_2+\frac{1}{\sqrt{2}}e^{i(\frac{\chi_1}{2}+\chi_2-\frac{\chi_3}{2})}\Ket{\widetilde{1}}_3\\
\\
\Ket{1}_3&\rightarrow&\frac{1}{\sqrt{2}}\Ket{\widetilde{0}}_2-\frac{1}{\sqrt{2}}e^{i(\frac{\chi_1}{2}+\chi_2-\frac{\chi_3}{2})}\Ket{\widetilde{1}}_3\\
\end{array}
\end{equation}
that returns the GHZ, and this ends the proof.

\bigskip

\section{Discussion}\label{sec:Discussion}

In this paper, by introducing a simple algebraic approach we have
reproduced some known results which provide recipes to establish
whether a certain multipartite pure state is entangled with
respect to a given bipartition. Such an analysis allows to
determine whether the pure state describing a system is completely
separable, separable or totally entangled. In the last case one
can say that such a state possesses genuine multipartite
entanglement.

Our treatment naturally led us to introduce a class of functionals
which include quantities traceable back to the concepts of purity
and linear entropy. Moreover, we have dealt with the problem of
the simultaneous maximization of the relevant functionals
(purities of the reduced matrices), providing an alternative proof
to the known result that in the case of three qubits the only
states that maximize all the relevant functionals are the
GHZ-state and all the equivalent states (i.e., equal up to local
unitary transformations).

This preliminary studies paves the way to the analysis of
multipartite entanglement in cases wherein more than three
subsystems are involved. In particular, simultaneous maximization
of the relevant functionals for $N$-partite systems --- with $N
\ge 4$ --- could provide interesting results and ideas in the
study of maximal multipartite entanglement. It is of relevance to
stress in addition that other known functionals might be found in
the class we have introduced, or, alternatively, new quantities
could be introduced and their properties explored.

\section{Acknowledgements}

The authors wish to thank D. Chru\'sci\'nski, H. de Guise, A.
Jamio{\l}kowski and M. Michalski for stimulating discussions. The
authors express their gratitude to P. Facchi for interesting
suggestions and for carefully reading the manuscript.

\bigskip



\begin{thebibliography}{99}

\bibitem{ref:Vedral1997} V. Vedral and M. B. Plenio, Phys. Rev. A
\textbf{57}, 1619 (1997).

\bibitem{ref:p96}
A. Peres, Phys. Rev. Lett. \textbf{77}, 1413 (1996).

\bibitem{ref:Horodecki1997}
P. Horodecki, Phys. Lett. A \textbf{232}, 333 (1997).

\bibitem{ref:ho}
R. Horodecki, P. Horodecki, M. Horodecki and K. Horodecki, Rev. Mod.
Phys. \textbf{81}, 865 (2009).

\bibitem{ref:hw} S. Hill and W. K. Wotters, Phys. Rev. Lett. \textbf{78}, 5022 (1997).

\bibitem{ref:w98}
W. K. Wootters, Phys. Rev. Lett. \textbf{80}, 2245 (1998).

\bibitem{ref:re}
L. Amico, R. Fazio, A. Osterloh and V. Vedral, Rev. Mod. Phys. \textbf{80}, 517
(2008).

\bibitem{ref:man}
V. I. Man'ko, G. Marmo, E. C. G. Sudarshan and F. Zaccaria, J. Phys. A: Math. Gen. \textbf{35}, 7137 (2002) .

\bibitem{ref:pf}
P. Facchi, Rend. Lincei Mat. Appl. \textbf{20}, 25-67 (2009).

\bibitem{ref:ga}
C. Sabin and G. Garcia-Alcaine, Eur. Phys. J. D. \textbf{48}, 435-442
(2008).

\bibitem{ref:fa}
F. Anz\`{a}, B. Militello and A. Messina, J. Phys. B \textbf{43},
205501 (2010).

\bibitem{ref:tt}
V. Coffman, J. Kundu and W. K. Wootters, Phys. Rev. A \textbf{61}, 052306
(2000).

\bibitem{ref:ttt}
E. Jung, D. Park and J. W. Son, Phys. Rev. A \textbf{80}, 010301 (2009).

\bibitem{ref:classification}
H. A. Carteret and A. Sudbery, J. Phys. A: Math. Gen. \textbf{33}, 4981 (2000).

\bibitem{ref:mm}
H. M\"{a}kel\"{a} and A. Messina, Phys. Rev. A \textbf{81}, 012326 (2010).

\bibitem{ref:hyp}
A. Miyake, Phys. Rev. A \textbf{67}, 012108 (2003).

\bibitem{ref:hu}
M. Huber, F. Mintert, A. Gabriel and B. C. Hiesmayar, Phys. Rev. Lett.
\textbf{104}, 210501 (2010).

\bibitem{ref:te}
B. Militello and A. Messina, Phys. Rev. A \textbf{83}, 042305 (2011).

\bibitem{ref:Fano1957} U. Fano, Rev. Mod. Phys. \textbf{29}, 74
(1957).

\bibitem{ref:Jaeger2003}
G. Jaeger, A. V. Sergienko, B. E. A. Saleh and M. C. Teich, Phys.
Rev. A \textbf{68}, 022318 (2003).

\bibitem{ref:osterloh}
S. J. Akhtarshenas, J. Phys. A: Math. Gen. \textbf{38} (2005) 6777–6784.

\bibitem{ref:huber1}
F. Mintert, M. Ku\'{s} and A. Buchleitner Phys. Rev. Lett. \textbf{92}, 167902 (2004).

\bibitem{ref:huber2}
F. Mintert, M. Ku\'{s} and A. Buchleitner Phys. Rev. Lett. \textbf{95}, 260502 (2005).

\bibitem{ref:Equiv_neg}
S. Shelly Sharma and N. K. Sharma, Phys. Rev. A \textbf{82}, 012340
(2010).

\bibitem{ref:GHZ}
J. Schlienz and G. Mahler, Phys. Lett. A \textbf{224} (1996) 39-44













\end{thebibliography}
\end{document}